# Foreshocks and b-value: bridging macroscopic observations to source mechanical considerations

## M. Avlonitis and G.A. Papadopoulos

Department of Informatics, Ionian University,Platia Tsirigoti 7, Corfu, Greece,

e-mail:avlon@ionio.gr

Institute of Geodynamics, National Observatory of Athens, Athens, Greece,

e-mail: papadop@noa.gr

**Abstract**

Spring-block models, such as the Olami-Feder-Christensen (OFC) model, were introduced several years ago to describe earthquake dynamics in the context of self-organized criticality. With the aim to address the dependency of the seismicity style on source's material properties we present an analytical enrichment of a 2D OFC model. We concluded with an analytical expression which introduces, through an appropriate constitutive equation, an effective dissipation parameter $a^{eff}$ related analytically not only with the elastic properties of the fault plane, but also with stochastic structural heterogeneities and structural processes of the source through a gradient coefficient. Moreover, within the proposed formulation the low $b$-values experimentally observed in foreshock sequences can be modeled by a process of material softening in the seismogenic volume. To check our analytical findings a cellular automaton was build-up whereas simulation results have verified model's predictions for the evolution of $b$ in macroscopic records.

## 1. Introduction

The forecasting or prediction of earthquakes remains an ever-green topic in the area of seismology in the last decades. In the aftermath of the lethal earthquake (Mw=6.3) of 6th April 2009 in L' Aquila, Italy, it was shown that a strong foreshock signal preceded the mainshock in the domains of space, time and size (Papadopoulos et al., 2010), and a discussion re-opened regarding the operational implementation of the forecasting of tectonic earthquakes (Jordan et at., 2011). Time independent forecasts are useful for long-term seismic hazard analysis. However, in a time-dependent forecast, the probabilities depend on the information available at time when the forecast is made. In the same work, it was found that the most useful information for operational forecasting has come from seismic catalogs and the geologic history of surface ruptures.

In the short-term sense the incidence of foreshocks as a precursory pattern useful for the forecasting of the mainshock attracted the interest of the seismological community since the 1960's (Mogi, 1963a,b). Since foreshocks may provide information which may lead to a more robust estimation of the probability for the occurrence of a future strong mainshock (e.g. Agnew and Jones, 1991), they are generally considered as one of the most promising precursory phenomena (Wyss, 1997, Vidale et al., 2001). However, some mainshocks are preceded by foreshocks while others are not, which underlines the need to investigate the physical conditions that favour or disfavour the foreshockoccurence.

The methodological approach introduced in this paper is to bridge mechanical considerations of the earthquake source to macroscopic observations resulting from seismicity analysis. In particular, we investigate links between foreshock activity and one of the most important parameters of seismicity, the "$b$-value" of the magnitude-frequency or G-R relation (Ishimoto and Iida, 1939, Gutenberg and Richter, 1944),

$$\log N = d - bM \tag{1}$$

where $N$ is the cumulative number of events of magnitude equal to or larger than $M$ and $d$, $b$ are parameters determined by the data. It is shown that relaxing the assumption of constant shear strength within the earthquake source it is possible to analytically model the effect of both the material heterogeneity and structural evolution of the source to the macroscopically observed $b$-values. More specifically, it is shown how the introduction of a softening mechanism before mainshocks can adequately model the macroscopically observed drop of b-value in foreshock sequences.

In the next two sections we review shortly macroscopic observations as well as the OFC spring-block model. Then, our contribution in understanding the variability of $b$-value via analytical derivation is presented, while the predicted behavior of $b$-variability is validated via simulation results.

## 2. Macroscopic observations

Seismicity space-time clusters have been examined around the world long ago. Omori (1894) was the first to introduce the concept of aftershock sequence, that is of a space-time cluster of shocks following a larger mainshock with power-law time decay of the number of aftershocks known as Omori's law. Later on, three patterns of earthquake sequences were described by Mogi (1963b): mainshock-aftershocks, foreshocks-mainshock-aftershocks, swarms. The mainshock-aftershocks pattern implies that foreshocks precede only some mainshocks and not others, although perhaps it is only an apparent result given that, in the routine seismic analysis performed in seismograph centers, low-magnitude shocks including short-term foreshocks  usually escape recognition and are not listed in standard earthquake catalogues due to limitations in monitoring capabilities. A detailed case study in the east Aegean Sea was published by Papadopoulos et al. (2006).  It is of relevance to note that the completeness magnitude cut-off in the earthquake catalogues varies from one region to another again due to different monitoring capabilities. Spatio-temporal seismicity clusters that exhibit a gradual rise and fall in seismic moment release, lacking a mainshock-aftershocks pattern, are termed earthquake swarms (Yamashita, 1998). It is quite common that in a swarm the magnitude range is too short which results in high or even very high values of $b$.

Foreshock sequences are characterized by some distinct features. Laboratory material fracture experiments (e.g. Mogi, 1963a, Scholz, 1968) along with numerical modeling in spring-block models (e.g. Hainzl et al., 1999) and analytical damage mechanics modeling (e.g. Main, 2000) showed a clear acceleration of the fracturing process before the main fracture. Studies regarding seismicity in Japan, western United States, Greece, Italy and elsewhere verified this in nature showing that foreshock activity increases approximately as the inverse of time before mainshock (e.g. Papazachos, 1975, Kagan and Knopoff, 1978, Jones and Molnar, 1979, Papadopoulos et al., 2000, 2010, 2011). In the G-R relation  and in global scale seismicity $b$ was found around unity (Frohlich and Davis, 1993). This parameter, however, is a variable dependent on local seismotectonic conditions, such as the material heterogeneity, the degree of symmetry in stress distribution, and on the existence or not of asperities in the fault zone. Such conditions are reflected in the mode of seismic activity, e.g. background seismicity, swarms, foreshocks, aftershocks.

Observations on seismic sequences have shown that the parameter $b$ usually drops and becomes significantly lower in foreshocks than in aftershocks or in background seismicity (Papazachos, 1975, Jones and Molnar, 1979, Main et al., 1989, Molchan et al., 1999, Papadopoulos et al., 2010, 2011). On the other hand, in laboratory experiments, analysis of seismicity in mines, pore pressure records and simulations in spring-block models indicated that locally high stress may drop the normal $b=1$ to lower values (Mogi, 1963b, Scholz, 1968, Wyss, 1973, Main et al., 1989, Urbancic et al., 1992, Hainzl et al., 1999). On the contrary, creeping segments of the fault are found to be characterized by high $b$-values (Amelung and King, 1997), which is consistent with the high $b$-value usually found in aftershock sequences as compared to the $b$-value in foreshocks and in the background seismicity. Therefore, for seismically active parts of the fault zone it was proposed that $b$ can be used as an indicator where asperities may be located, that is where the creeping and locked patches are situated along the fault zone, and where mainshocks are most likely (Wyss et al., 2000, Wyss, 2001, Zhao and Wu, 2008). Swarm-type activity is characterized by high to very high values of the parameter $b$ at around 1.5 or more. The strong variations of $b$ across different stress regimes implies that this parameter acts as a stress meter that depends inversely on differential stress (Schorlemmer et al., 2005, Narteau et al., 2009).

From the point of view of mechanical properties of the earthquake source, there is evidence that the incidence of foreshocks is disfavored by increasing focal depth and that likely depends on the faulting type and orientation (Jones, 1984, Ohnaka, 1992, Abercrombie and Mori, 1996, Maeda, 1996, Reasenberg, 1999). Given that normal stress in the crust, that is the regional tectonic stress plus the loading stress, increases with depth, supports the hypothesis that increasing normal stress inhibits foreshock occurrence and that at high normal stress foreshock sequences tend to be short or non-existent (Abercrombie and Mori, 1996). Since heterogeneity decreases with depth and normal stress, foreshock activity should increase with the degree of small-scale crustal heterogeneity. The stronger the heterogeneity the larger the percentage of mainshocks preceded by foreshock sequences and the longer the duration of each individual foreshock sequence.

### 3.  Source considerations and the OFC model

Most models of earthquake dynamics have in common that if the applied stress between the two fault segments exceeds a certain threshold value a displacement occurs instantaneously and an earthquake is generated. In order to model this kind of elastic behavior, the so-called spring-block models were introduced such as the BK (Burridge and Knopoff 1967) and OFC (Olami et al. 1992) models. Although a wide variety of spring-block models were proposed and studied by several authors (see review in Rundle et al., 2003), so far they are only weakly related to distributed seismicity (Turcotte et al., 2009). A modification of the OFC model was studied by Jagla (Jagla 2010) who introduced a structural relaxation mechanism, thus generating earthquake sequences that contain features observed in real seismicity. The relaxation was introduced in a rather phenomenological way by means of an arbitrary energy measure $E$ stored locally in the system and which can not directly mapped to some known source mechanical parameters.

In this paper we address the mapping of source mechanical parameters to macroscopic observations by presenting a theoretical framework  which models analytically the dependence of the macroscopic b-values on the structural heterogeneity as well as on material processes, such as structural relaxation, through  appropriate constitutive equations. A specific case of the theoretical framework is examined where a model for structural softening is proposed by relaxing the assumption of constant shear strength in order to reproduce realistic b-values in foreshocks sequences.

Following the lines of the OFC model we note that at the basis of spring-block models which exhibit Self-Organized Criticality (SOC) we find the assumption of two rigid plates, moving relative each other, where their interface is modeled by means of a number of blocks (nodes) interconnected with springs of a certain elastic constants, *K*, for each direction. If a one-dimension scheme is adopted the well-known BK model is obtained. Here we consider the two-dimension scheme of the problem as it was presented by the OFC model depicted in Fig. (1).

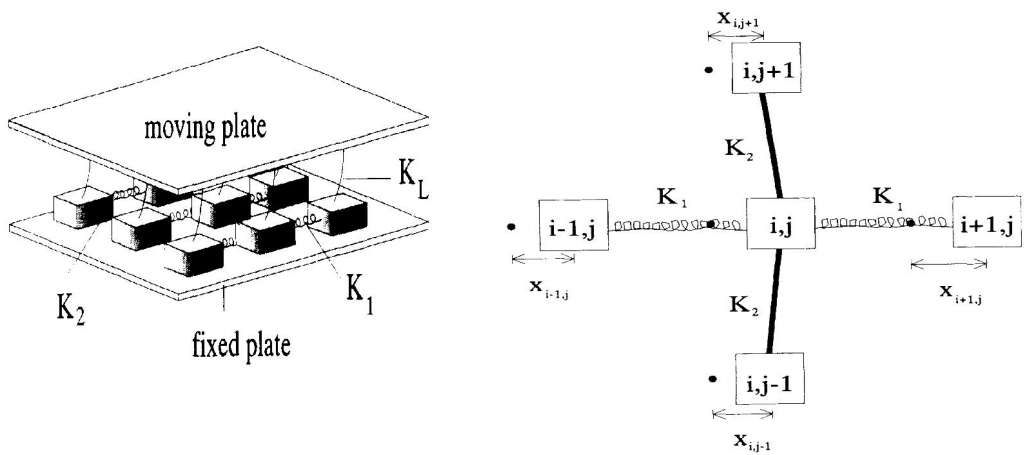

Figure 1. The geometry of the OFC model for earthquake source (after Olami et al. 1992).

A dynamic variable, $f_{ij}$, is assigned to each node of a two-dimension grid which represents the fault plane of an earthquake source. The variable, $f_{ij}$, being the total force exerted at the node $n_{ij}$, grows through time, *t*, at a constant quantity, $r = K_L \upsilon t$ where $\upsilon$ is the constant velocity of the upper moving plate and $K_L$ is the elastic constant at the vertical direction of the fault plane. Assuming without loss of generality the isotropic case $K_1 \equiv K_2 \equiv K$ at the horizontal plane ($K_1, K_2$ are the elastic constants along the two axes), then the dynamic variable, in the continuum limit, is expressed by ($u_{ij}$ being the node's displacement at $x_{ij}$)

$$f_{ij} = Kl^2 \nabla^2 u_{ij} + K_L \delta u_{ij} \qquad (2)$$

where $\delta u_{ij} = \upsilon t - u_{ij}$ and the second order gradient term in the right hand-side models as usual in the continuum limit the expression $2u_{ij} - u_{i-1j} - u_{i+1j} + 2u_{ij} - u_{ij-1} - u_{ij+1}$. $l$ is a characteristic scale associated with space discretization (hereafter, without loss of generality, $l$ is assumed unity). When the force on an arbitrary node is larger than a threshold value $F_{th}$, which is the maximal static friction, the node slips. For the node $n_{ij}$ of the fault plane to become unstable the following inequality must be fulfilled,

$$K\nabla^2 u_{ij} + K_L \delta u_{ij} - F_{th} \geq 0 \qquad (3)$$

Then, redistribution of forces to the neighbor sites is taking place as follows,

$$f_{ij} \rightarrow 0, \ f_{mm} \rightarrow a f_{ij} \qquad (4)$$

where $f_{mm}$ are the forces for the four nearest neighbors and $a$ is the ratio at which force is transferred to the neighbor sites. For the assumed isotropic case the parameter $a$ takes the form $a = 1/(4 + K_L / K)$ (Olami et al. 1992) and can be interpreted as a parameter of the source's material depending on the corresponding elastic constants of the fault plane. Note that since always $K_L / K > 0$ then always $a < 0.25$ and as a result the OFC model describes a non-conservative system. Decreasing $a$ reflects increase in dissipation, while when this parameter equals 0.25 the case is the conservative one.

Within this context, simulations of the original OFC model by means of a continuous automaton qualitatively predicted the emergence of power-law for the earthquake magnitude distribution. On the other hand, initially the presence of foreshocks and aftershocks was not recognized in the OFC model as pointed out, for example, by Hainzl et al. (1999). However, Hergarten and Neugebauer (2000) and Hergarten (2002) showed with the help of numerical simulations that the OFC model exhibits foreshocks and aftershocks with results consistent with Omori's empirical law for the time decay of events, although the exponents predicted by the model are lower than observed in nature. Then, foreshocks and aftershocks can be attributed to the non-conservative character of OFC model. On the other hand, the experimentally observed variations of $b$-value before and after the occurrence of an earthquake has not been modeled so far in terms of SOC.

From simulation results, Olami et al. (1992) showed that in their model the G-R power-law, and consequently the $b$-value, depends reversely on the conservation level. The measured values of $b$ around 1 are obtained for $a \approx 0.20$. With the decrease of $a$ the critical exponent $b$ increases. Furthermore, it was found that the exponent varies continuously when changing the boundary conditions smoothly (Christensen and Olami, 1992). These results indicated that there is no universality of the critical exponent. Introducing anisotropy changes the scaling of the distribution function, but not the power-law exponent. The dependence of the exponent $b$ on the conservation level, that is on $\alpha$, was also shown for simulated foreshock and aftershock sequences (Hergarten and Neugebauer, 2002).

At this point the contradiction of the OFC model with experimental findings arises: while the OFC model assumes a constant material parameter $a$ and consequently a constant power-law exponent $b$, the last varies significantly with the style of seismicity, e.g. before and after a mainshock or during seismic swarms. As a consequence the OFC model is not capable to correctly model the material processes taking place within the source, which seems to be more complex than described by the simplified picture of the spring-block model suggested earlier. In the next lines, we present an analytical enrichment of the spring-block model, and support it by simulation results, with the aim to address the dependency of the seismicity style on source's mechanical properties.

### 4. Analytical OFC modeling of the *b*-value variability

It has been proposed that the weaknesses of the existing models of SOC for seismicity lie in the fact that internal mesoscopic material processes, such as viscous relaxation, were not taken into account (e.g. Hainzl et al., 1999, Pelletier, 2000). Consequently, interactions between network sites are by no means the only dynamic process taking place but also material parameters, that were assumed constant so far, are also involved in the process of seismogenesis in the time and space domains. More recently, it was hypothesized that the threshold value $F_{th}$ of the maximal static friction is not a constant material parameter but a dynamic internal variable which varies with the evolution of seismicity (Avlonitis and Tassos, 2010). As will be shown in the next paragraphs, this leads to a drastic change in the dynamic behavior of the spring-block model for earthquakes, resulting to robust model predictions for the variability of the *b*-value.

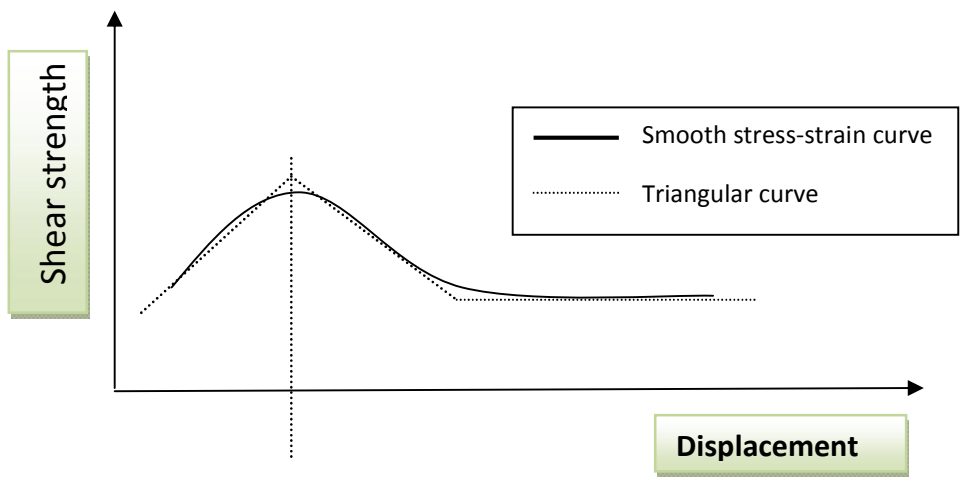

**Figure 2**. A non-convex shear strength hardening-softening law of the fault plane.

Let us assume a shear displacement induced hardening-softening law for the shear strength of the fault plane as proposed by Palmer and Rice (1973) and McClung (1979)

(Fig. 2). In the simplest case, adapting a linear idealization of the hardening-softening curve, we may write,

$$F_{th} = F_{th}(u_{ij}) = F^* + \lambda u_{ij} \qquad (5)$$

where $\lambda$ is a proportionality parameter modeling the hardening or softening process of the Earth's material and $F^*$ a constant parameter. The sign of the proportionality parameter (as well as the parameter $F^*$) depends at which branch of the hardening-softening law of the fault plane the material deforms. Within the framework of the so-called gradient theory, the non-local behavior of the displacement manifold can be modeled by gradient terms (Aifantis 2003, and references therein), i.e.,

$$u_{ij} = u_{ij}^0 + c\nabla^2 u_{ij} \qquad (6)$$

where $u_{ij}^0$ is a uniform over the entire space displacement field and $c$ is the gradient coefficient. The second order gradient term in the right hand-side of Eq. (6) was introduced in order to model the local material non-uniformities into the macroscopic stress-strain relation of Eq. (3). Substituting in (5), we get

$$F_{th} = F^* + \lambda(u_{ij}^0 + c\nabla^2 u_{ij}) \qquad \text{or} \qquad F_{th} = F_{th}^0 + K'\nabla^2 u_{ij} \qquad (7)$$

with $F_{th}^0 = F^* + \lambda u_{ij}^0$ and

$$K' = \lambda c \qquad (8)$$

After relaxing the assumption of constant homogeneous shear strength and interpreting it as an internal spatially evolving variable, the final inequality for the node $n_{ij}$ of the fault plane to become unstable, reads

$$(K - K')\nabla^2 u_{ij} + K_L \delta u_{ij} - F_{th}^0 \geq 0 \qquad (9)$$

As a result an effective dissipation parameter $a^{eff}$, which controls the macroscopically observed values of $b$, is introduced as follows, ,

$$a^{eff} = \frac{1}{4 + \dfrac{K_L}{K - K'}} \qquad (10)$$

Taking into account Eq. (8) we end up with the final expression

$$a^{eff}(c, \lambda) = \frac{1}{4 + \dfrac{K_L}{K - K'(c, \lambda)}} \qquad (11)$$

Eq. (11) is of crucial importance since it redresses the generic weakness of the OFC model. In fact, it relates the effective structural parameter $a^{eff}$ not only with $K, K_L$, that is the elastic properties of the fault plane, but also with structural heterogeneity through the gradient coefficient $c$. More importantly, Eq. (11) relates $a^{eff}$ with material processes of the earthquake source through an appropriate constitutive equation for source evolution as expressed in Eq. (5), i.e. whether the source material deforms under a softening or hardening mode expressed by the appropriate sign -minus or plus correspondingly- of the proportionality parameter $\lambda$.

Within the proposed physical framework, a one-to-one relation between an effective structural parameter $a^{eff}$ and the macroscopically observed $b$-value is provided either theoretically or synthetically through simulation experiments on the enriched OFC model. As a result, within this context it is possible to bridge the macroscopically observed $b$-values with specific styles of seismicity and, therefore, to specific processes of seismogenesis. Then, the low $b$-values experimentally observed in foreshock sequences can be modeled or mapped by a process of material softening in the seismogenic volume. Indeed, as the source material enters to the descending branch in Fig. (2), Eq. (5) takes the following form (for convenience the term $F^*$ was omitted),

$$F_{th} = F_{th}(u_{ij}) = -|\lambda| u_{ij} \quad (12)$$

The negative value $-|\lambda|$, results to negative values of $K'(c,\lambda)$ and, via Eq. (11), as $|\lambda|$ increases the effective structural parameter $a^{eff}(c,\lambda)$ increases too. However, from simulation experiments it comes out that $a^{eff}$ and $b$ evolve always reversely as noted before. Then, as the source material enters to the softening regime macroscopic lower values of $b$ are expected, which is in perfect accordance with the macroscopic observations in foreshock sequences. As a consequence, within the proposed analytical formulation a bridge between macroscopic observations and specific material processes in the source was established, providing a robust tool in the effort to understand better the behavior of seismicity not only in the presence of foreshocks at the final acceleration stage before mainshock generation, but also in aftershocks, in swarms and in the background seismicity as well.

## 5. Simulation results

To check the analytical findings of our formulation presented before a cellular automaton was build-up in order to simulate the enriched OFC model. Practically, if the simulation of the evolution of the source material when enters to the softening regime is the goal, then a variation of the classic OFC simulator is needed. Then, we can reconstruct the macroscopic magnitude-frequency relation and check whether lower *b*-values are emerged in accordance to observations in foreshock sequences. To this end, in order to mimic the evolution of source material when enters to the softening regime we built the classical OFC simulator as explained in Section 3 and introduced a linear reduction of the threshold value $F_{th}$ of each site according to Eq. (5). The results presented below refer to the values $a^{OFC} = 0.245$ and $a^{OFC} = 0.235$ of the OFC classical structural parameter but it is noted that similar behavior was found to lower values of $a^{OFC}$ for which more realistic values of *b* are reproduced.

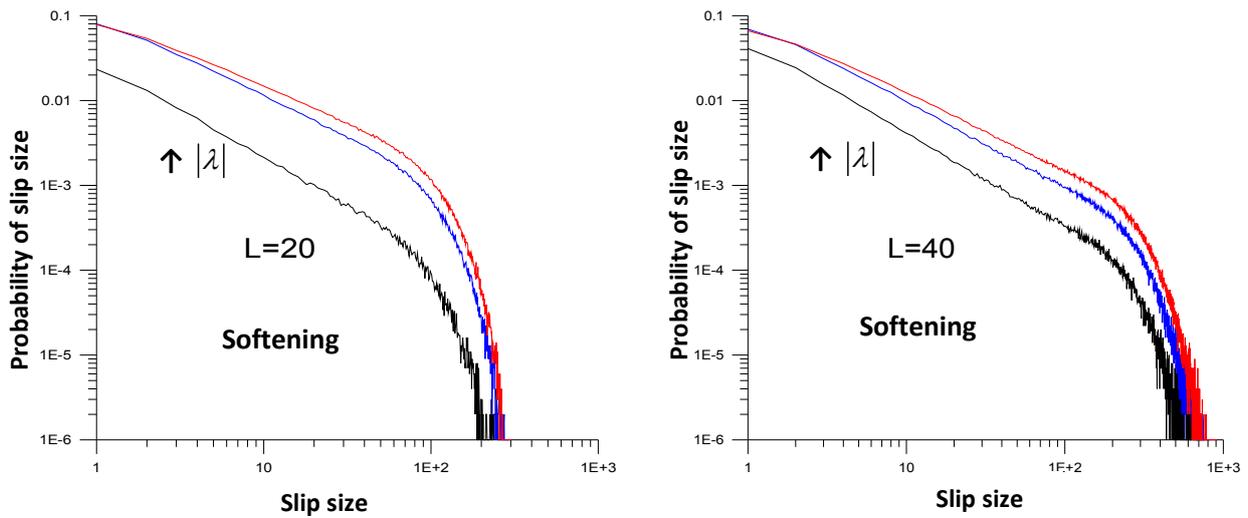

**Figure 3**. Simulation results for the macroscopic magnitude-frequency relation within the softening regime for different system sizes, L=20, 40 and for $a^{OFC} = 0.245$. Cases of no softening (first curve at the bottom) and gradually increasing softening are shown.

Our main finding consists of the variation of the *b*-value with parameter $\lambda$ of the introduced material process, that is hardening or softening. Indeed, in Fig. (3) the cases of no softening and of gradually increasing softening are depicted for different system sizes indicating that both the qualitative and quantitative behavior of the system dynamics are similar for different system sizes. Moreover, it is evident that the parameter $b$ changes inversely proportionally with the softening coefficient, in accordance with our findings in the previous sections. The procedure is analytically presented in Figure 4. For different values of the softening parameter, $\lambda$, mainly when $\lambda = 0, -2, -4$, the estimation of the emerged $b$-value is given.

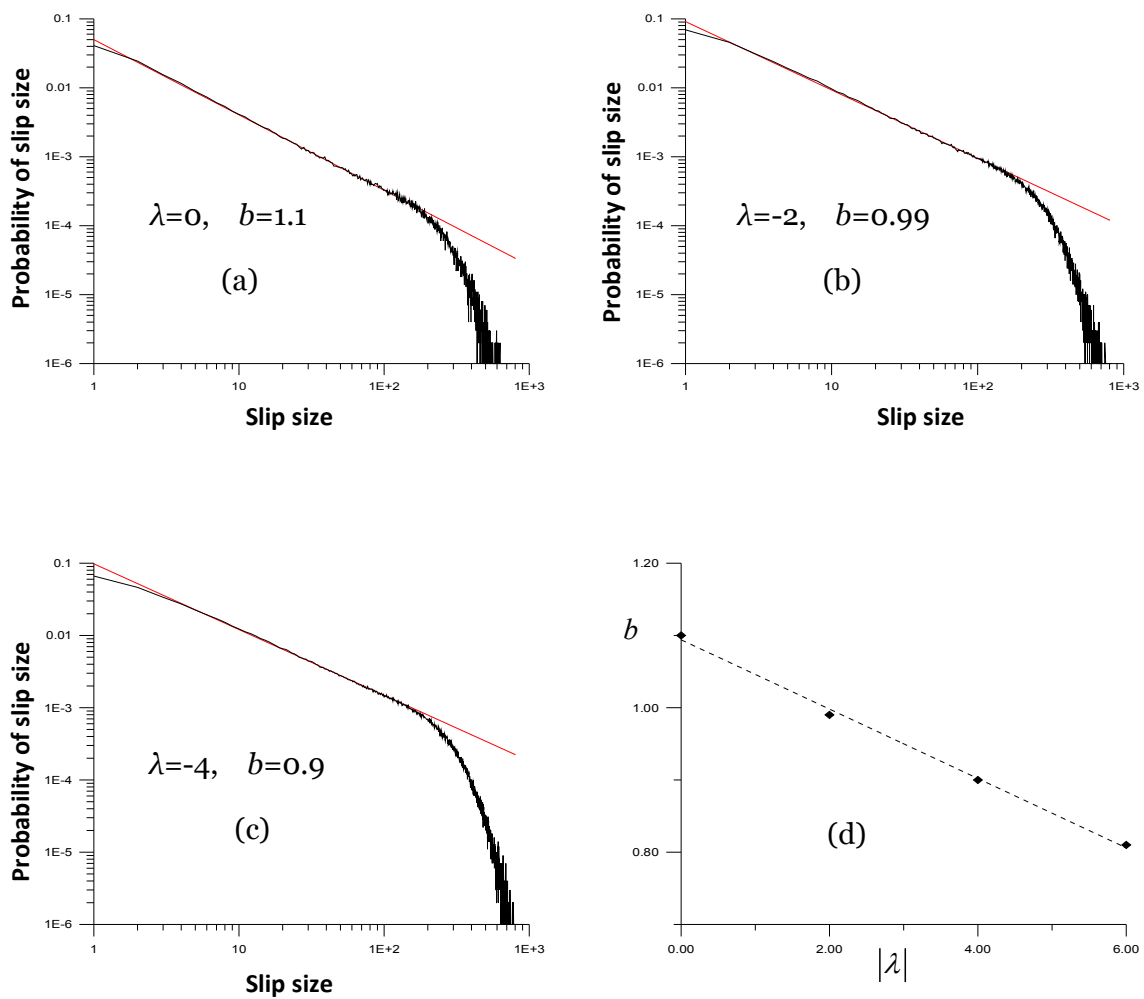

**Figure 4**. Explicit estimation of the b-values within the softening regime (a), (b), (c). In (d) a clear linear relation between $b$ and $|\lambda|$ is depicted. The simulations referred to L=40 and $a^{OFC} = 0.245$.

It can be seen that a clear linear reduction of the *b*-value versus the absolute value of the softening parameter $\lambda$ is emerged (Fig. 4d).

Although the linear dependency of the macroscopic *b*-value on the softening coefficient within the softening regime is pronounced this is not the case within the hardening regime, that is for positive $\lambda$. In fact, while an initially significant increase for the corresponding *b*-exponent is found, an almost no dependency regime is emerged for greater hardening parameters (Fig. 5). Notably the dynamics breaks down for hardening parameters greater than unity.

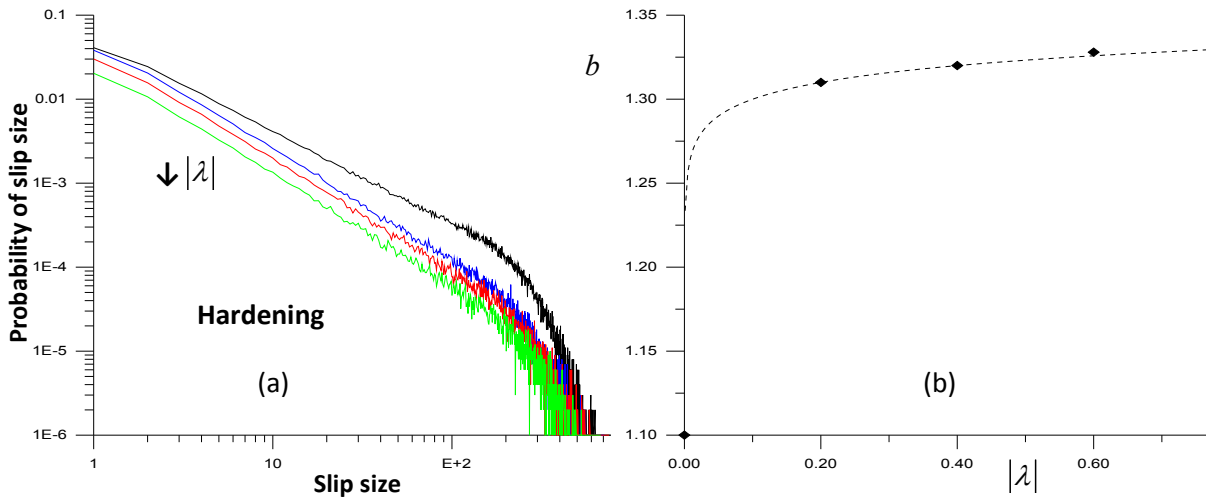

**Figure 5**. In (a) simulation results for the macroscopic magnitude-frequency relation within the hardening regime is shown for system size L=40 and for $a^{OFC} = 0.245$. In (b) the almost no dependency regime between $b$ and $|\lambda|$ is depicted.

In Fig. 6a a comparison of the estimated linear relation of the *b*-values versus the softening parameter $|\lambda|$ is depicted for different values of the structural parameter $a^{OFC}$. It can be seen that while the linearity is preserved a slightly decrease of the corresponding slope is observed. Finally in Fig. 6(b) the dependency of the *b*-value from the structural heterogeneity is also depicted. Here we mimic source heterogeneities by introducing a distribution of the shear strength values. Indeed, the different curves correspond to different Weibull distributions of the shear strength, $F_{th}$. More specifically, different simulations were performed for different shape parameters $w$ of the Weibull distribution ( $p(x;m,w) = (w/m)(x/m)^{w-1} \exp\left[-(x/m)^w\right]$, $x \geq 0$ ) of $F_{th}$, thus imposing gradually stronger heterogeneity: higher $w$ corresponds to higher variability in the corresponding distribution and as a result to higher heterogeneity. It can be seen that the stronger the heterogeneity the higher the $b$-value, in accordance with our derivation in Eq. (10) and because, as we noted before, the dissipation parameter $a^{eff}$ and the $b$-value always evolve reversely.

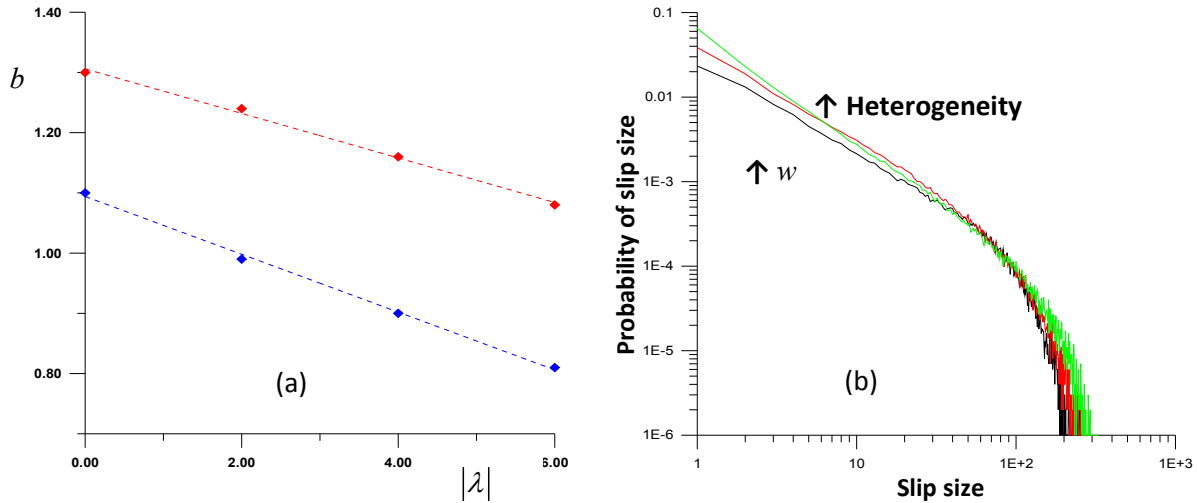

**Figure 6**. (a) Comparison of the estimated linear relation of $b$ vs $|\lambda|$ for different values of the structural parameter, $a^{OFC} = 0.245$ and $a^{OFC} = 0.235$, and for L=40. (b) Probability distributions for different levels of heterogeneity of $F_{th}$ by means of different shape parameters $w$ of the Weibull distribution.

**Discussion**

The simulation results presented in the previous section refer to an artificial situation where the earthquake source evolves always under softening. This simplified simulation setup should be elaborated by introducing a material mechanism that reproduces the non-convex hardening-softening law of the fault plane, as illustrated in Figure (2). As an instance, it could be the mechanism of transient creep as was proposed by Hainzl et al. (1999). In any case, the adaptation of the specific mechanism taking place at the source will be incorporated into the equation that introduces the effective parameter $a^{\mathit{eff}}$ as expressed here by Eq. (11). Moreover, the dependency of the $b$-value from the structural heterogeneity through the gradient coefficient $c$, as expressed in Eq. (11), implies that different structures at the source may result in different macroscopic $b$-values. This became evident from our simulation results (Fig. 6b). According to Aki (1981), if there is a basic generic difference between foreshocks and normal earthquakes (background seismicity), there remains a hope to discriminate them. Smaller $b$-values observed for some foreshocks than for aftershocks certainly agree with the idea that the percolation model applies to the former and the barrier model to the latter (Aki, 1981).

However, what certainly remains unexplained is why some mainshocks are preceded by foreshocks and others are not. This behaviour may be related to mainshocks that take place under conditions of plasticity. It has been proposed that when a mainshock nucleates within the brittle seismogenic layer and near the base of the seismogenic layer, immediate foreshocks for this mainshock are necessarily restricted to lie within a localized region shallower than hypocentral depth of mainshock (Ohnaka, 1992). By contrast, the nucleation process below the base of the seismogenic layer is aseismic in nature, as it happens with interplate earthquakes that nucleate below the base of the brittle seismogenic layer, thus explaining also why strong intermediate-depth and deep earthquakes are not associated by foreshocks while the aftershock activity is much lower than in shallow earthquakes (see review in Frolich, 2006). Roberts and Turcotte (2000) considered the plastic instability hypothesis against mechanical friction, provided numerical solution using plastic rheology and concluded that because the yield (failure) stress decreases with increasing temperature, earthquake nucleation at the base of the seismogenic zone follows naturally. Plastic rheology includes work hardening and thermal softening. In this context the proposed formulation could be a valuable tool in order to model such material processes.

As a conclusion, the dependency of the seismicity style, e.g. foreshocks or aftershocks, on source's mechanical properties was studied by an analytical enrichment of a 2D spring-block model of OFC type. The analytical expression reached relates, through an appropriate gradient term, the newly introduced generalized effective dissipation parameter $a^{eff}$ not only with the elastic properties of the fault plane, but also with structural heterogeneity and specific structural processes of the source. Simulations results with a modified OFC cellular automaton have verified the predicted evolution of $b$ in macroscopic records, i.e., low $b$-values the experimentally observed in foreshock sequences can be modeled by a process of material softening in the seismogenic volume that is by a process of strain rate increasing with time under constant stress. During aftershocks and beyond, however, it is not expected that the material processes affect drastically the macroscopic behavior as expressed by the magnitude-frequency relation.